\documentclass[twocolumn,10pt,prl,nofootinbib]{revtex4}
\usepackage[utf8]{inputenc}

\def\mysections#1{{\bf #1.} }
\usepackage{hyperref}
\usepackage{cleveref}
\usepackage[dvips]{graphicx}
\usepackage{epsfig,amsmath,amssymb,verbatim,mathrsfs,array,layout,textcomp,amssymb,latexsym,slashed,graphicx,booktabs,color,mathtools,tikz}
\usepackage{tikz-feynman}
\newcommand*{\Scale}[2][4]{\scalebox{#1}{$#2$}}%

\newcommand{\beq}{\begin{eqnarray}}% can be used as {equation} or {eqnarray}
\newcommand{\eeq}{\end{eqnarray}}

\def\beqa{\begin{eqnarray}}
\def\eeqa{\end{eqnarray}}
\newcommand{\no}{\nonumber}
\newcommand{\bv}{\left(\begin{array}{c}}
\newcommand{\ev}{\end{array}\right)}
\newcommand{\bmtwo}{\left(\begin{array}{cc}}
\newcommand{\bmthree}{\left(\begin{array}{ccc}}
\newcommand{\emn}{\end{array}\right)}
\newcommand{\bmtwoc}{\left\{\begin{array}{cc}}
\newcommand{\bmthreec}{\left\{\begin{array}{ccc}}
\newcommand{\emnc}{\end{array}\right\}}
\newcommand{\ba}{\begin{array}}
\newcommand{\ea}{\end{array}}

\def\lsim{\mathrel{\rlap{\lower4pt\hbox{\hskip1pt$\sim$}}
     \raise1pt\hbox{$<$}}}         %less than or approx. symbol
\def\gsim{\mathrel{\rlap{\lower4pt\hbox{\hskip1pt$\sim$}}
     \raise1pt\hbox{$>$}}}         %greater than or approx. symbol

\begin{document}
%\vspace*{-30mm}
\font\mini=cmr10 at 0.8pt

\title{
Dark Spectroscopy
}

\author{Yonit Hochberg${}^{1,2}$}\email{yonit.hochberg@cornell.edu}
\author{Eric Kuflik${}^{1,2}$}\email{kuflik@cornell.edu}
\author{Hitoshi Murayama${}^{3,4,5}$}\email{hitoshi@berkeley.edu, hitoshi.murayama@ipmu.jp}
\affiliation{${}^1$Department of Physics, LEPP, Cornell University, Ithaca NY 14853, USA}
\affiliation{${}^2$Racah Institute of Physics, Hebrew University of Jerusalem, Jerusalem 91904, Israel}
\affiliation{${}^3$Ernest Orlando Lawrence Berkeley National Laboratory, University of California, Berkeley, CA 94720, USA}
\affiliation{${}^4$Department of Physics, University of California, Berkeley, CA 94720, USA}
\affiliation{${}^5$Kavli Institute for the Physics and Mathematics of the
  Universe (WPI), University of Tokyo Institutes for Advanced Study, University of Tokyo,
  Kashiwa 277-8583, Japan}

%\date{\today}
\begin{abstract}
  Rich and complex dark sectors are abundant in particle physics theories. Here we propose performing spectroscopy of the mass structure of  dark sectors via
  mono-photon searches at lepton colliders. The energy of the mono-photon tracks the invariant mass of the invisible system it recoils against,
  which enables studying the resonance structure of the dark sector. We demonstrate this idea with several well-motivated models of dark sectors.
  Such spectroscopy measurements could potentially be performed at Belle II, BES-III and future low-energy lepton colliders. %and at a future tau-charm collider.
\end{abstract}

\maketitle

%%%%%%%%%%%%%%%
\section{Introduction}

The existence of dark matter (DM) is by now well-established, though its exact identity is unknown. Theoretical proposals for its particle nature
span many orders of magnitude in mass, with various possible mechanisms for setting its relic abundance. Often, DM is part of a larger dark
sector, comprised of a wealth of resonances, and can exhibit rich dynamics. Moreover, complex dark sectors can arise in models beyond the Standard
Model (SM), irrespective of candidates for dark matter. The possibility to experimentally study the structure of dark sectors is therefore an
extremely relevant and important task.

In the context of dark sectors, expansive attention has been devoted to dark photons that are kinetically mixed with the SM hypercharge~\cite{Holdom:1985ag}.
Constraints from beam dumps, fixed-target experiments, B-factories, stellar environments and colliders have been widely studied in the literature
(see {\it e.g.} Refs.~\cite{Bergsma:1985is,Konaka:1986cb,Riordan:1987aw,Bjorken:1988as,Bross:1989mp,Davier:1989wz,Athanassopoulos:1997er,Astier:2001ck,Adler:2004hp,Bjorken:2009mm,Artamonov:2009sz,Essig:2010gu,Blumlein:2011mv,Gninenko:2012eq,Blumlein:2013cua,Abrahamyan:2011gv,Merkel:2014avp,Merkel:2011ze,Aubert:2009cp,Curtin:2013fra,Lees:2014xha,Bernardi:1985ny,MeijerDrees:1992kd,Archilli:2011zc,Gninenko:2011uv,Babusci:2012cr,Adlarson:2013eza,Agakishiev:2013fwl,Adare:2014mgk,Batley:2015lha,Anastasi:2016ktq,Pospelov:2008zw,Chang:2016ntp}, and Ref.~\cite{Alexander:2016aln} for a recent review of this topic). Colliders such as LEP, BaBar, and the LHC have access to the light
states of a dark sector by
searching for events with missing energy.
  Such searches are generically sensitive to dark sector states, but do not directly probe the spectrum of the dark sector, which can often be rich with dark meson resonances.

\begin{figure}[t!]
\tikzfeynmanset{large}
\begin{tikzpicture} \begin{feynman}
/tikzfeynman/large
\vertex (a1) {\(\Scale[1.5]{ e^-}\)};
    \vertex[right=2cm of a1] (a2);
    \vertex[right=2cm of a2] (a3) {\(\Scale[1.5]{\gamma}\)};

\vertex[below=2cm of a1] (b1) {\(\Scale[1.5]{ e^+}\)};
    \vertex[right=2cm of b1] (b2);
     \vertex[right=1cm of b2] (b0);
     \vertex[above=0.1cm of b0] (bV){\(\Scale[1.5]{V}\)}  ;
    \vertex[right=1cm of b0] (b3) ;
     \vertex[right=1 cm of b3] (b4);
	\vertex[above=.25 cm of b4] (b5) ;
 	\vertex[below=.25 cm of b4] (b6) ;
  	\vertex[above=.5 cm of b4] (b7) ;
  	  	\vertex[below=.5 cm of b4] (b8) ;         	
\diagram* {
      {[edges=fermion]
        (a1) -- (a2) -- (b2) -- (b1),
      },
       (b3) -- [boson] (b2),
       (a2) -- [boson, momentum=\(E_\gamma\)] (a3),
       (b3) -- [plain] (b4),
 {[edges=plain]
        (b6) -- (b3) -- (b5),
      },
  {[edges=plain]
        (b7) --  (b3) -- [edge label'=\( \,\,{\rm invisible}\)]  (b8),
      },
    };

    	\vertex[right=.1 cm of b7] (b9) ;
  	  	\vertex[right=.1 cm of b8] (b10) ;
    	\vertex[above=.05 cm of b9] (b11) ;
  	  	\vertex[below=.05 cm of b10] (b12) ;     	  	
 \draw [decoration={brace}, decorate] (b11.north east) -- (b12.south east)
          node [pos=0.5, right] {\({M_{\rm inv}}\)};

\end{feynman} \end{tikzpicture}
\caption{\label{diagram} Mono-photon production at a lepton collider.}
\end{figure}
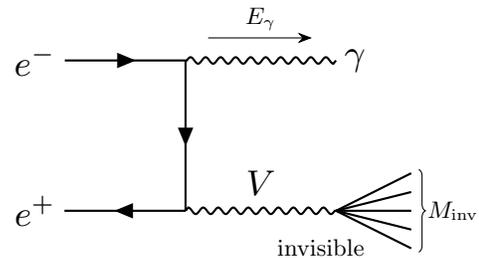

Here we propose a method to probe the resonance spectrum of a dark sector. The idea is simple, and we first draw an analogy from QCD. At an $e^+e^-$ machine, the resonance structure of QCD can be mapped by scanning the center of mass energy of the collision.
Similarly, the resonance spectrum can be studied by looking at $e^+e^-\to \gamma+{\rm hadrons}$ events at a fixed center of mass energy
collision. There, the mono-photon energy traces the mass of the system it recoils against, thus performing spectroscopy even at fixed center of mass energy. As the observation of the resonances is not required, such a measurement can easily be performed on a dark sector, where the resonances may be invisible. A schematic description of this proposed {\it dark spectroscopy} is given in Fig.~\ref{diagram}.

We propose, and study the feasibility of, performing spectroscopy of generic dark sectors at low-energy colliders. The specific case of a Strongly Interacting Massive Particle (SIMP) dark sector~\cite{Hochberg:2014dra,Hochberg:2014kqa}, was previously considered by the authors in Ref.~\cite{Hochberg:2015vrg}.
%  Performing spectroscopy on the resonance structure of the dark sector for strongly coupled sectors was studied in Ref.~\cite{Hochberg:2015vrg} in the context of Strongly Interacting Massive Particle (SIMP) dark matter~\cite{Hochberg:2014dra,Hochberg:2014kqa}.
(For enhanced signals from dark matter bound states in a weakly coupled sector, see Ref.~\cite{An:2015pva}.) However, the concept outlined in Ref.~\cite{Hochberg:2015vrg} applies more broadly to any strongly coupled dark sector that interacts with the SM, regardless of the nature of dark matter within the model. It is the purpose of this paper to demonstrate this in a concrete manner.

%%%%%%%%%%%%%%%
\section{Concept}

\begin{figure*}[t!]
\begin{center}
	\includegraphics[width=0.48\textwidth]{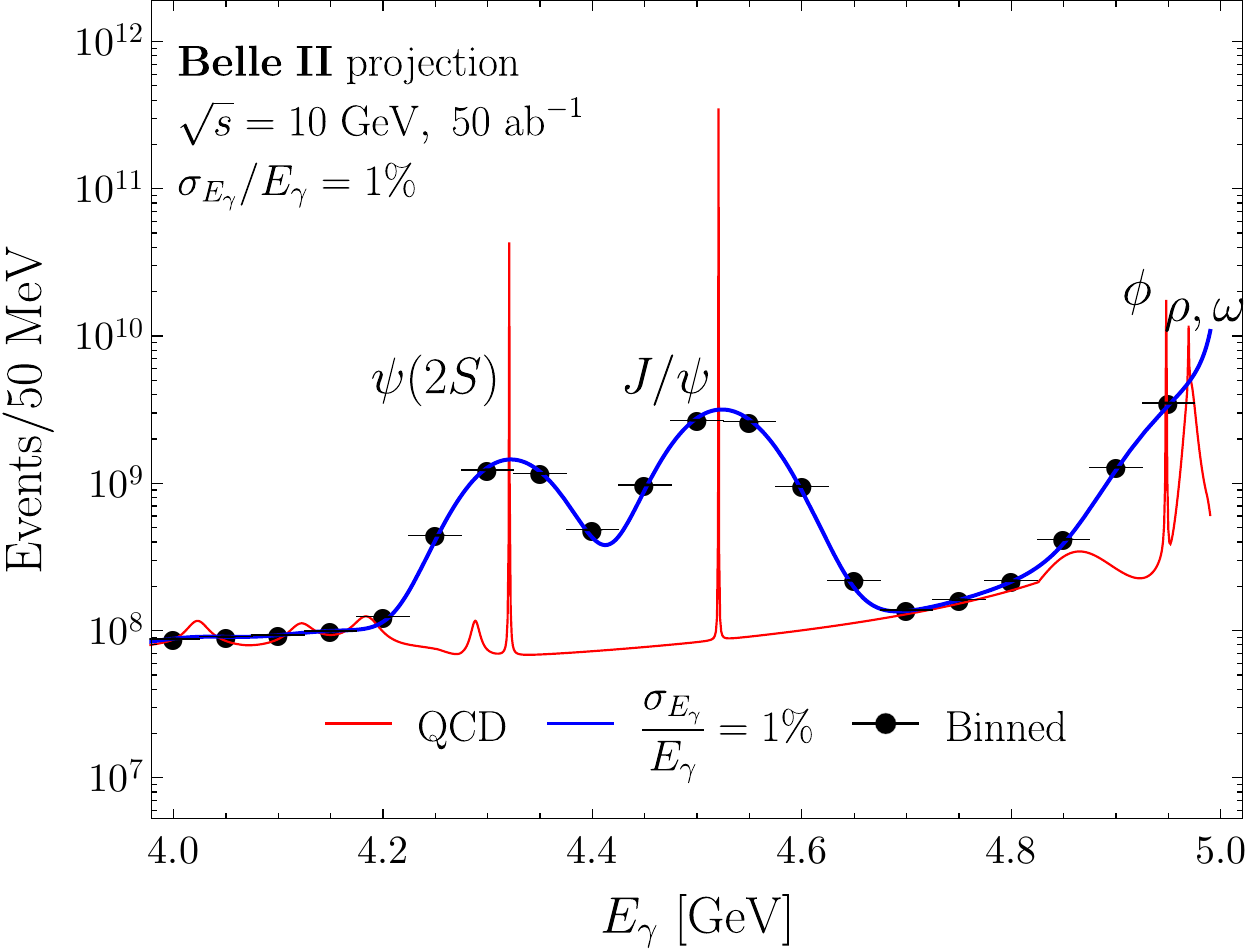}~\hfill \includegraphics[width=0.48\textwidth]{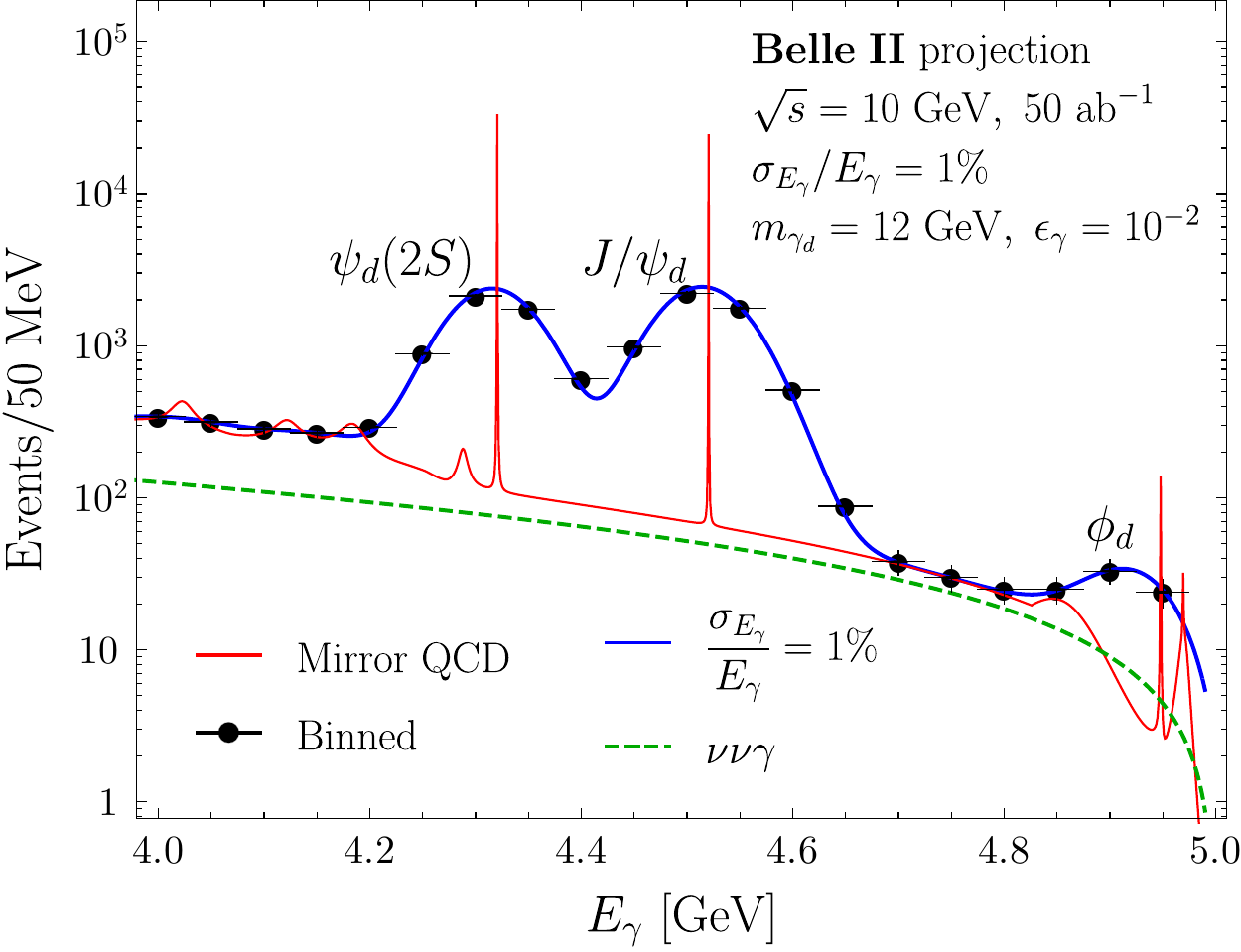}
	\end{center}
	\vspace{-0.3 cm}
 \caption{{\bf Left}: Distribution for $e^+e^- \to \gamma + had$ for QCD, when ignoring the hadronic activity, at $\sqrt{s}=10$~GeV
 at Belle II with 50~ab$^{-1}$ of data (red), and the expected results for 1\% energy resolution (blue) and binned data (black). {\bf Right}: The expected distribution for $e^+e^- \to \gamma + {\rm inv}$ for a mirror copy of QCD, with $m_{\gamma_d}=12$~GeV and $\epsilon_\gamma=10^{-2}$, for  50~ab$^{-1}$ at Belle II with  $\sqrt{s}=10$~GeV (red), 1\% energy resolution (blue), and binned data (black). The SM background of $e^+e^- \to \gamma\nu\nu$ is shown in green in both panels.
  \label{fig:QCD}}
\end{figure*}

In a $\ell^+\ell^-\to \gamma + {\rm inv}$ collision at center-of-mass energy $\sqrt{s}$, the energy of the outgoing photon is in one-to-one
correspondence with the invariant mass of the invisible system, $M_{\rm inv}$, it recoils against:
\beq
E_\gamma=\frac{\sqrt{s}}{2}\left(1-\frac{M_{\rm inv}^2}{s}\right)\,.
\eeq
Thus by measuring the photon energy, one can determine the spectrum of the undetected system in the process.

For concreteness, we study the case in which a dark sector communicates with the visible sector via a vector, $V$. In later sections we will take $V$ to be dark $U(1)_d$ photon $\gamma_d$, which is
kinetically mixed with hypercharge. The mono-photon production cross section at a lepton collider can then be written as
\beq\label{eq:sigmono}
&&\sigma(\ell^+\ell^-\to \gamma+{\rm inv}) =\\
&& \frac{3\alpha}{s}\int d\; {\rm cos}\theta \int d M_{\rm inv}^2\frac{M_{\rm inv}^2}{(M_{\rm inv}^2-m_V^2)^2+m_V^2 \Gamma^2}\times \no\\
&&\frac{\Gamma_{V\to {\rm inv}}(M_{\rm inv})}{M_{\rm inv}}\frac{\Gamma_{V\to e^+e^-}(M_{\rm inv})}{M_{\rm inv}}\frac{8-8\beta+3\beta^2+\beta^2{\rm
cos}2\theta}{\beta{\rm sin}^2\theta}\no\,,
\eeq
where $\beta=1-M_{\rm inv}^2/s$ and the decay widths are to be computed for $m_V=M_{\rm inv}$, reflecting the off-shell nature of the vector in the process.

The irreducible SM background of $\ell^+\ell^- \to \gamma \nu\bar\nu$ which proceeds via an off-shell $Z$ is easily obtained from
Eq.~\eqref{eq:sigmono} by taking $m_V= m_Z$ and replacing the invisible width by the $Z$-width into neutrinos. (Note that the contribution from $W$-fusion is negligible.) Additional backgrounds arise from
$\ell^+\ell^-\to \gamma\gamma$ (peaked at $M_{\rm inv}^2=0$), as well as from $\ell^+\ell^-\to \gamma \gamma \gamma$ and $\ell^+\ell^-\to\gamma
\ell^+\ell^-$  when only one photon is observed due to other particles going undetected down the beam-pipe or in a detector
crack. Such backgrounds can potentially be mitigated with knowledge over the location of the detector cracks.

We study the potential of low-energy electron colliders,
such as Belle II and BES-III, to probe the spectroscopy of
the dark sector. Belle II is expected to operate at $\sqrt{s}=10$~GeV with 50~ab$^{-1}$ of data and ${\cal O}(1-2\%)$ energy resolution at large $E_\gamma$~\cite{Higuchi}. BES-III operates at lower center-of-mass-energy $\sqrt{s}\sim 4$~GeV, with anticipated 10~fb$^{-1}$ and $\sim2\%$ energy resolution at high photon energies~ \cite{Briere:2016puj}.
To demonstrate the potential reach of these machines, photon energies are smeared using a Gaussian distribution with a given energy resolution. We
will take the photon acceptance to be $|\cos\theta|<\cos12^\circ$, motivated by the geometric coverage of Belle II, between 12 to 157~degrees.
%When considering binned distributions, we use a constant percentage energy resolution with a fixed bin size equal to the energy resolution for the highest mono-photon energy.

%%%%%%%%%%%%%%%
\section{Results}

\subsection{Standard Model QCD}
\begin{figure*}[t!]
\begin{center}
    \includegraphics[width=1.01\textwidth]{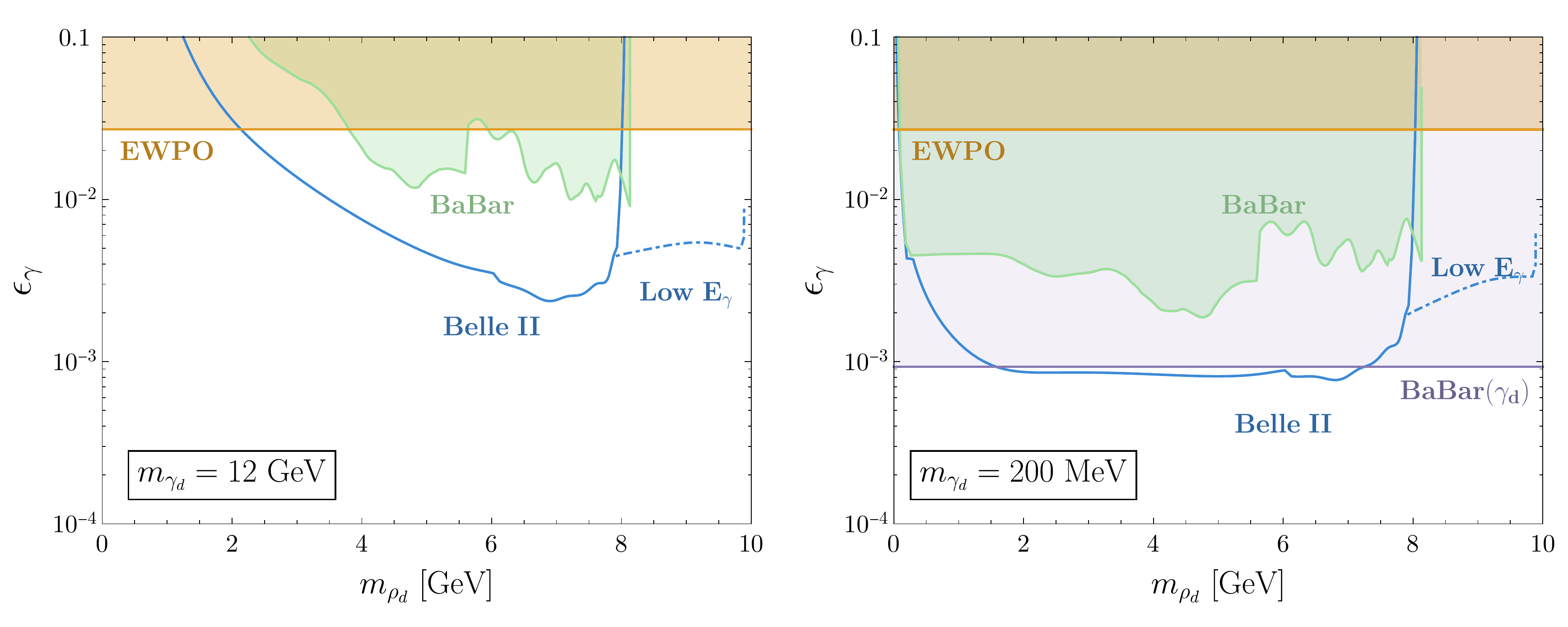}\hfill
\end{center}
\vspace{-0.5 cm}
 \caption{Constraints on the dark $\rho_d$ parameter space, from electroweak precision observables~\cite{Hook:2010tw} (shaded orange), BaBar search for dark photons~\cite{Lees:2017lec} (shaded purple), translated BaBar constraints for $\rho_d$~\cite{Lees:2017lec} (shaded green) and the estimated reach at Belle II~\cite{Essig:2013vha} (solid and dashed blue curves), for a heavy ({\bf left}) and light ({\bf right}) dark photon.
  \label{fig:reach}}
\end{figure*}

We begin by examining what the resonances of QCD would look like in mono-photon events when ignoring the hadronics in the process, as if they had decayed `invisibly'.    In the SM, the cross section for the lower resonances are dominated by off-shell photons.\footnote{A low-energy description of the vector meson couplings to leptons,  called vector-meson dominance,  can be given by an effective meson kinetic-mixing with the photons. In this description, the process proceeds via on off-shell photon that mixes with the hadronic resonance. We will use this to describe production of dark-resonances, via an off-shell massive vector, below.}
The width of the off-shell photon can be found from the total hadronic cross-section in $e^+ e^-$ annihilations. Using standard notation, the total cross-section at center of mass energy $\sqrt{s}$ is
\begin{equation}\label{eq:Rs}
R(s) \equiv \frac{\sigma(e^+ e^- \rightarrow
\mbox{hadron})}{\sigma(e^+ e^- \rightarrow \mu^+ \mu^-)_0}\,,
\end{equation}
where the subscript $0$ refers to the lowest order QED calculation for massless muons,
\beq
\sigma(e^+ e^- \rightarrow \mu^+ \mu^-)_0 = \frac{4\pi\alpha^2}{3s}\,.
\eeq
The data for $R(s)$ is taken from Refs.~\cite{Ezhela:2003pp,Agashe:2014kda}. By cutting the photon propagator in the diagram, $R(s)$ can be written in terms of the the off-shell widths at $m_{\gamma^*} = \sqrt{s}$,
\beq
R(s) = \frac{\Gamma_{\gamma^*\rightarrow
\mbox{hadrons}}({\sqrt{s}})}{\Gamma_{\gamma^* \rightarrow \mu^+ \mu^-}({\sqrt{s}})}\,.
\eeq
The off-shell hadronic width at $M_{\rm inv}$ to be used in Eq.~(\ref{eq:sigmono}) is
\begin{equation}\label{eq:GammaQCD}
  \Gamma_{\gamma^* \rightarrow \rm had}(M_{\rm inv})
  =
    R(M_{\rm inv}) \frac{\alpha M_{\rm inv}}{3}\,,
\end{equation}
and the off-shell electron width,
\begin{equation}\label{eq:Gammall}
  \Gamma_{\gamma^* \rightarrow  e^+ e^-}(M_{\rm inv})
  = \frac{\alpha M_{\rm inv}}{3}\,.
\end{equation}

The resulting spectrum for $\sqrt{s}=10$~GeV at Belle II with 50~ab$^{-1}$ is shown in the left panel of Fig.~\ref{fig:QCD} as a function of the mono-photon energy. We show the QCD result in red, compared to the smeared cross section given a $1\%$ energy resolution, shown in blue, as well as the number of events in 50~MeV bins, with Poisson variation. The low-mass resonances show up as one, with the dominant contribution coming from the $\phi$. The higher mass $J/\psi$ and $\psi(2S)$ resonances are clearly visible in the distribution, though the energy resolution of the machine significantly broadens them.

%------------ Dark QCD ------------------------------

%
\subsection{Mirror QCD}

Next we consider a mirror copy of SM, with the low-energy resonance structure of the mirror QCD sector identical to that of QCD. Such a scenario can be motivated by mirror models~\cite{Foot:1991bp,Foot:1991py,Berezhiani:1995am} or models of neutral naturalness~\cite{Chacko:2005pe,Burdman:2006tz,Cai:2008au,Craig:2014aea,Craig:2014roa,Chacko:2005vw,Batell:2015aha,Arkani-Hamed:2016rle}. We assume that there is kinetic mixing between the SM hypercharge and mirror photon, labeled $B_\mu$ and $\mathcal{A}_\mu$ respectively, however with a massive mirror photon,
\beq
\mathcal{L} = \mathcal{L}_{\rm SM} + \mathcal{L}_{\rm Mirror} + \frac{1}{2} m_{\gamma_d}^2 \mathcal{A}_{ \mu}  \mathcal{A}^{\mu} - \frac{\sin\chi}{2} B_{\mu \nu }  \mathcal{A}^{\mu \nu}\,. \label{eq:kinmix}
\eeq

Here the production of dark mesons proceeds through an additional off-shell mirror photon via the kinetic mixing to the off-shell SM photon. The widths in Eqs.~\eqref{eq:GammaQCD} and ~\eqref{eq:Gammall} are modified to the case at hand (see e.g. Ref.~\cite{Batell:2009yf}):
\beqa
  \Gamma_{\gamma_d^* \rightarrow \rm inv}(M_{\rm inv})&=& \frac{\alpha \epsilon_\gamma^2}{\alpha_D}R(M_{\rm inv}) \frac{\alpha_D M_{\rm inv}}{3}\,,\nonumber\\
    \Gamma_{\gamma_d^* \rightarrow  e^+ e^-}(M_{\rm inv})
  &=& \frac{\alpha \epsilon_\gamma^2}{\alpha_D}\frac{\alpha_D M_{\rm inv}}{3}\,,
\eeqa
where  $\epsilon_\gamma \simeq - c_w\sin\chi$.
For a full list of conventions, see Section~3.1 of Ref.~\cite{Hochberg:2015vrg}.

We take the kinetically mixed dark photon mass and kinetic mixing parameter to obey all existing constraints (see {\it e.g.} Ref.~\cite{Hochberg:2015vrg}), and show the expected binned distribution at Belle II in right panel of Fig.~\ref{fig:QCD}, for  $m_{\gamma_d}=12$~GeV and $\epsilon_\gamma=10^{-2}$, along with the SM background.
The heavy dark quarkonium resonances $\psi_d(2S)$ and $J/\psi_d$ are clearly visible in the mono-photon distribution, as is the presence of lower resonances around the dark $\phi_d$. The narrow states provide a large enhancement in the signal relative to the perturbative prediction if one neglects the resonance structure: for Belle II, the total number of events around the $J/\psi_d$~[$\psi_d(2S)$] peak is around $~\sim10$~$[\sim16]$ times larger than the perturbative continuum. A measurement of the cross-section will allow for determination of the dark quark masses and strong coupling constant at the scale of the resonance.   We learn that Belle II could shine light on the structure of a mirror QCD.

\subsection{Dark Sectors}

\begin{figure}[t!]
\begin{center}
	\includegraphics[width=0.48\textwidth]{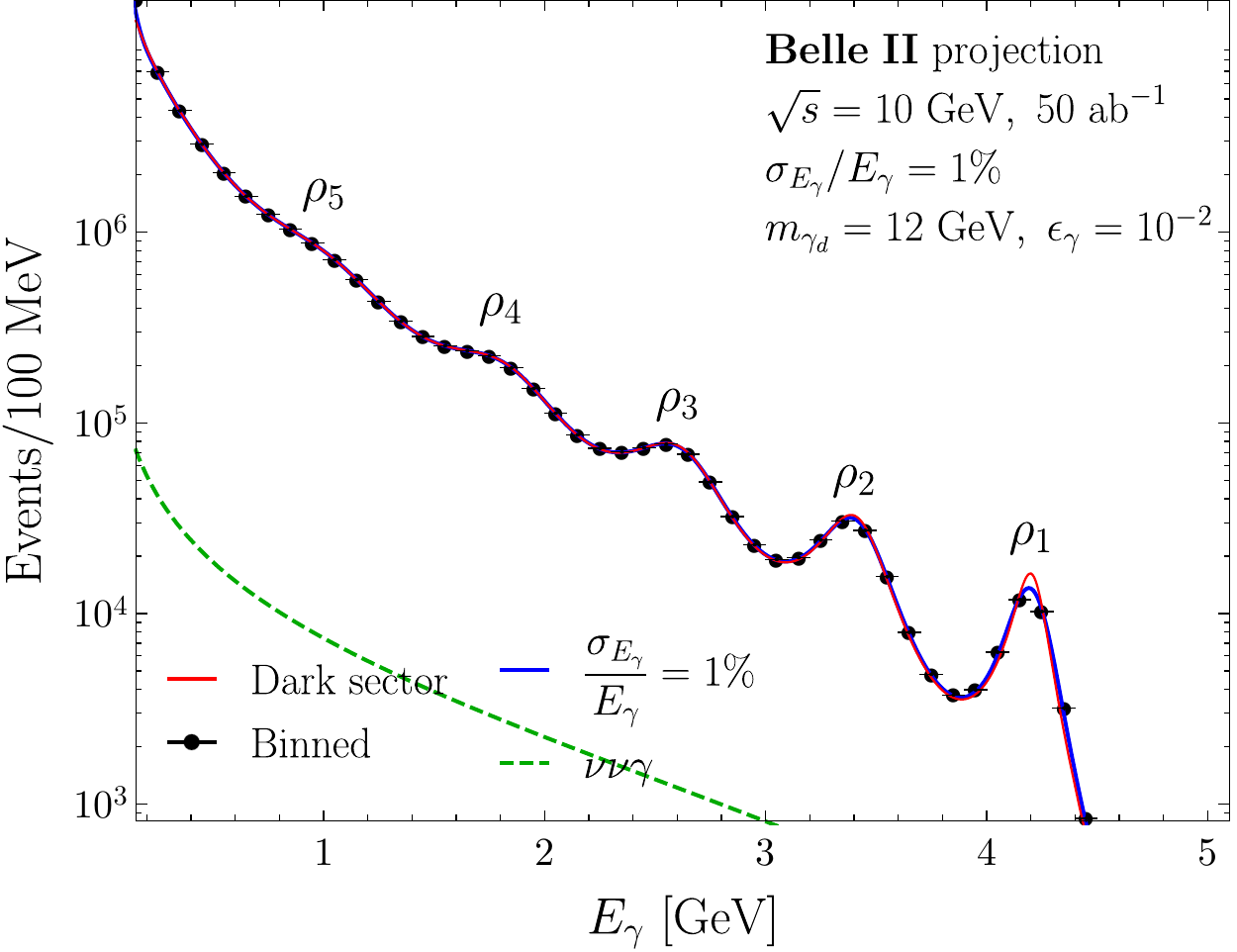}
\end{center}
 \caption{The cross section for $e^+e^- \to \gamma + {\rm inv}$ at $\sqrt{s}=10$~GeV
 for an $SU(2)_d\times U(1)_d$ gauge theory with 4 Weyl fermions, $m_{\gamma_d}=12$~GeV, $m_{\pi_d}=1$~GeV, $m_{\rho_1}=4$~GeV, $\epsilon_\gamma=10^{-2}$ and $\alpha_D=0.1$, with $1\%$ energy resolution. The SM background $e^+e^-\to \gamma \nu\nu$ is shown in green.
  \label{fig:SU2heavy}}
\end{figure}

\begin{figure*}[t!]
\begin{center}
    \includegraphics[width=\textwidth]{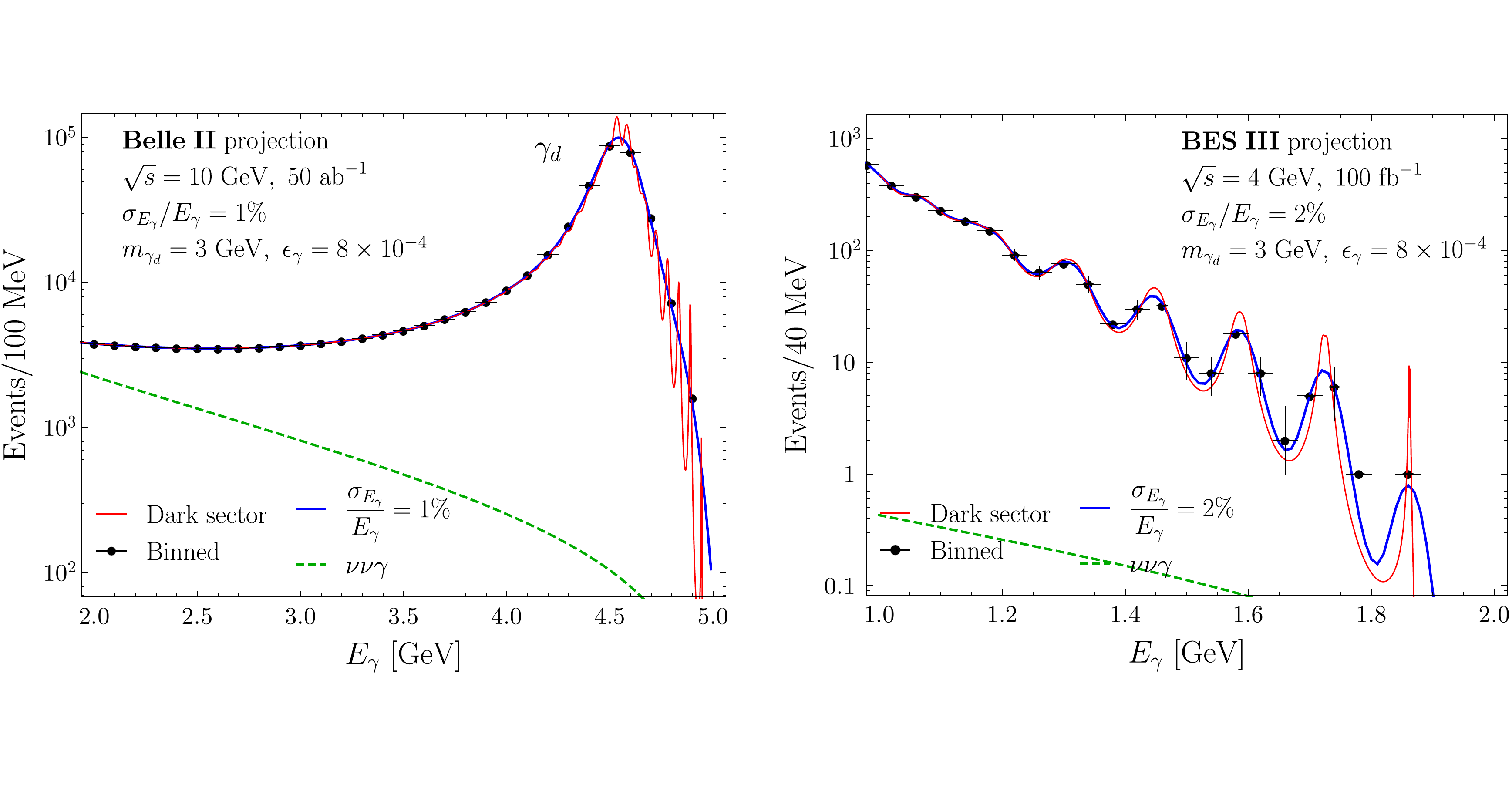}
\end{center}
\vspace{-1.5cm}
 \caption{The cross section for $e^+e^- \to \gamma + {\rm inv}$
 for SIMP dark matter of an $Sp(4)_d\times U(1)_d$ gauge theory with 4 Weyl fermions, $m_{\gamma_d}=3$~GeV, $m_{\pi_d}=500$~MeV, $m_{\rho_1}/m_{\pi_d}=2.1$, $\epsilon_\gamma=8\times10^{-4}$ and $\alpha_d=1/(4\pi)$, for Belle II at $\sqrt{s}=10$~GeV with 50~ab$^{-1}$ of data ({\bf left}) and for BES-III with high luminosity of 100~fb$^{-1}$ of data ({\bf right}), with 1\% and 2\% energy resolution, respectively.
 The SM background of $e^+e^-\to \gamma\nu\nu$ is shown in green in both panels.
  \label{fig:SIMP}}
\end{figure*}

Next we consider examples of the resonance structure of generic dark sectors such as strongly coupled theories inspired by the models of SIMP dark matter~\cite{Hochberg:2014dra,Hochberg:2014kqa}.
In general, we will consider confining gauge theories with a gauged dark  $U(1)_d$ which kinetically mixes with hyperchage via the Lagrangian Eq.~\eqref{eq:kinmix}.  Then production of singlet vector mesons (singlets under both the flavor symmetry and $U(1)_d$) will proceed via kinetic mixing with the dark photon $\gamma_d$. We will refer collectively to the pseudo Nambu-Goldstone bosons as dark pions $\pi_d$, and to the light singlet vector mesons that strongly decay as dark rho-mesons $\rho_d$.

For the $\rho_d$ resonance spectrum, we use the partial widths modeled in Ref.~\cite{Hochberg:2015vrg} as inspired by
soft-wall QCD; for simplicity we summarize the relevant results here. Using the effective meson dominance Lagrangian of $\rho_d$-$\gamma_d$ mixing,
\begin{eqnarray}
    {\cal L}_{\rm int} =
 \frac{F_{\rho_n} e_D}{2m_{\rho_n}^2}  {\cal A}_{\mu\nu}^a \rho_n^{\mu\nu a}+ (e_D \mathcal{A}^a_\mu + g_\rho \mathcal{\rho}^a_\mu) J^{a\,\mu }\,,\label{vmd}
\end{eqnarray}
the spectrum and decay constants are given by
\begin{eqnarray}
  m_{\rho_n}^2 = n m_{\rho_1}^2 ,\quad  \frac{F_{\rho_n}^2}{m_{\rho_n}^2} = \frac{m_{\rho_1}^2}{2 g_{\rho}^2},\quad g_{\rho}^2 = \frac{12\pi^2}{N_c}, 
  \end{eqnarray}
with the $\rho \to \pi_d \pi_d $ partial widths of
\begin{equation}
  \Gamma_n^\rho =\frac{D_R}{4} \frac{\beta^3}{96\pi} g_{\rho }^2 m_{\rho_n}\,,
  \qquad \beta=\sqrt{1-\frac{4m_{\pi_d}^2}{m_{\rho_n}^2}}\, ,
\end{equation}
assuming all the dark pions $\pi_d$ are degenerate. Here, the factor $D_R= {\rm Tr}(T_{\rho})^2$ and $T_{\rho}$ is the  generator corresponding to the $\rho_d$-vector.

{\it Constraints and reach.}
Next, we  estimate the range of parameters in which dark-$\rho$ spectroscopy can be performed. The effective interaction Lagrangian Eq.~\eqref{vmd}  introduces effective $\rho_d$-$\gamma$ mixing of the size
\begin{equation}
\mathcal{L}_{\rm eff} = \frac{\epsilon_{\rho_n}}{2} { B}_{\mu\nu}^a \rho_n^{\mu\nu a},~~~   \epsilon_{\rho_n} \simeq \begin{cases}
 	\frac{F_{\rho_n} m_{\rho_n}^2 e_D}{ m_{\gamma_d}^2}  \epsilon_\gamma & m_{\gamma_d}  \gg m_{\rho_n}\\
 		\frac{F_{\rho_n} e_D}{ m_{\rho_n}^2}  \epsilon_\gamma & m_{\gamma_d} \ll m_{\rho_n}\\
  \end{cases}
\end{equation}
One can then translate the constraints and estimated reach of a  dark photon with mass and kinetic-mixing $\{m_{\gamma_d},\epsilon_\gamma\}$ onto a $\rho_d$ meson with mass and kinetic mixing $\{m_{\rho_n},\epsilon_{\rho_n}\}$.
These are show in Fig.~\ref{fig:reach} for high ({\it left}) and low ({\it right}) $m_{\gamma_d}$, for an $SU(2)_d\times U(1)_d$ gauge theory, with 2 massless quarks of charge 1, and $\alpha_D = 0.1$. We show the constraints from electroweak precision observables (EWPO)~\cite{Hook:2010tw} and the BaBar search for dark photons~\cite{Lees:2017lec}, which are independent of the dark sector dynamics and couplings. We also show the translated constraints for $\rho_d$ at BaBar~\cite{Lees:2017lec}, as well as the conservative projections of  Ref.~\cite{Essig:2013vha} for Belle II, which assumes a 1.7\% energy resolution. We note that the expected reach improves with $N_f,N_c$ and $\alpha_D$. Next, in order to resolve the different $\rho_d$ peaks from each other, the separation of the peaks need to be larger than the resolution, which imposes $m_{\rho_1} \gtrsim \sqrt{s \frac{\sigma_{E_\gamma}}{E_\gamma}}$. At Belle II with optimized 1\% energy resolution, this requires $m_{\rho_d} \gtrsim 1$~GeV. Finally, the $\rho_d$ bumps should be visible above the continuum dark hadronic production, which is achieved whenever the signal is visible.

{\it Generic dark sector. }
To exemplify the potential spectroscopy of a generic dark sector, we consider  an $SU(2)_d\times U(1)_d$ gauge theory with $N_f=2$ as above.
 In Fig.~\ref{fig:SU2heavy} we show the expected mono-photon energy distribution for  1\% energy resolution, using the expected
luminosity of 50~ab$^{-1}$ at Belle II, for the case $m_{\pi_d}=1$~GeV,  $m_{\rho_1}=4$~GeV and $\epsilon_\gamma = 10^{-2}$.  The spectrum is clearly visible at Belle II.

{\it SIMPs. }
Next, we consider a similar symmetry breaking pattern but with a different spectrum. Motivated by SIMP dark matter~\cite{Hochberg:2014dra,Hochberg:2014kqa,Hochberg:2015vrg}, where the relic abudance is controlled by $3\to2$ annihilations (and see Refs.~\cite{asher1,asher2} for contributions of semi-annihilations as well), we take an $Sp(4)_d$ gauge theory with 4 Weyl fermions, which after confinement gives rise to dark pions $\pi_d$ which can play the role of dark matter. We use $m_{\pi_d}=500$~MeV, $m_{\rho_1}/m_{\pi_d}=2.1$, $m_{\gamma_d}=3$~GeV, $\epsilon_\gamma=8\times10^{-4}$ and $\alpha_d=1/(4\pi)$, and show the resulting invariant mass distribution for $e^+e^-$ collisions for $\sqrt{s}=10$~GeV at Belle II
in the left panel of Fig.~\ref{fig:SIMP}. .
In this case, the $\rho_d$-resonances cannot be resolved at Belle II, but the dark-photon peak is clearly visible.
For comparison, we show the distribution for $e^+e^-$ collisions for lower $\sqrt{s}=4$~GeV at BES-III (with increased luminosity) in the right panel of Fig.~\ref{fig:SIMP}, where the resonances are visible. There can also be a kinematic sharp edge at $E_\gamma=1.875$~GeV, corresponding to $M_{\rm inv} = 2 m_{\pi_d}$. We learn that low energy lepton colliders such as BES provide complementary tools to higher energy machines such as Belle II in performing spectroscopy of dark sectors.

\section{Summary}

In this letter, we have proposed a method to study the spectrum of dark sectors via the measurement of mono-photon events at low energy lepton colliders, such as Belle II and BES. By considering well-motivated dark sectors, we have shown that such dark spectroscopy can successfully be performed at Belle II, BES and future lepton colliders, providing an novel new avenue in which to explore the riches of the dark world.

\mysections{Acknowledgments}
We thank Leor Kuflik for inspiration, and Maxim Perelstein for comments on the draft.
The work of YH is supported by the U.S. National Science Foundation, grant NSF-PHY-1419008, the LHC Theory Initiative. EK is supported by the NSF
under Grant No. PHY-1316222 and the Bethe Postdoctoral Fellowship.
HM was supported by the U.S. DOE under Contract DE-AC02-05CH11231, and
by the NSF under grants PHY-1316783 and PHY-1638509.  HM was also
supported by the JSPS Grant-in-Aid for Scientific Research (C)
(No.~26400241 and 17K05409), MEXT Grant-in-Aid for Scientific Research
on Innovative Areas (No. 15H05887, 15K21733), and by WPI, MEXT, Japan.

\bibliography{bibliospectroscopy}{}

\end{document}